\documentstyle [twocolumn,epsf]{mn}
\newcommand{\mincir}{\raise
-3.truept\hbox{\rlap{\hbox{$\sim$}}\raise4.truept\hbox{$<$}\ }}
\newcommand{\magcir}{\raise
-3.truept\hbox{\rlap{\hbox{$\sim$}}\raise4.truept\hbox{$>$}\ }}
\newcommand{\minmag}{\raise
-3.truept\hbox{\rlap{\hbox{$<$}}\raise5.truept\hbox{$<$}\ }}
\newcommand{\be}{\begin{equation}}
\newcommand{\ee}{\end{equation}}
\newcommand{\ba}{\begin{eqnarray}}
\newcommand{\ea}{\end{eqnarray}}
\newcommand{\brr}{\begin{array}}
\newcommand{\err}{\end{array}}
\newcommand{\bc}{\begin{center}}
\newcommand{\ec}{\end{center}}

\title[ROSAT observations of ELAIS sources]
{The European Large Area ISO Survey VII: ROSAT observations of ELAIS sources}
\author[Basilakos, S. et al.]{S. Basilakos$^{1}$, I. Georgantopoulos$^{2}$, I. P\'erez-Fournon$^{3, 4}$, 
A. Efstathiou$^{1}$, 
\newauthor
M. Rowan-Robinson$^{1}$, F. Cabrera-Guerra$^{3}$, E. Gon\'zalez-Solares$^{3}$,
\newauthor
D. M. Alexander$^{5}$, S. Serjeant$^{6}$ and S. Oliver$^{7}$  \\
\vspace{0.1cm}
$^1$ Astrophysics Group, Imperial College London, Blackett Laboratory, 
Prince Consort Road, London SW7 2BW, UK\\
$^{2}$ Institute of Astronomy and Astrophysics, National Observatory of Athens, Palaia Penteli, 15236, Athens, Greece\\
$^{3}$ Instituto de Astrofisica de Canarias, Via Lactea s/n, 38200 La Laguna, Tenerife, Spain\\
$^{4}$ Instituto de Astrofisica de Canarias, Universidad La Laguna, Tenerife, Spain\\
$^{5}$ Department of Astronomy and Astrophysics, 525 Davey Laboratory, Pennsylvania State University, University
Park, PA 16802, USA \\
$^{6}$ Unit for Space Sciences and Astrophysics Scoolo of Physical Sciences, University of Kent, Canterbury Kent, 
CT2 7NR, UK\\
$^{7}$ Astronomy Centre, CPES, University of Sussex, Falmer, Brighton BN1 9QJ, UK\\ 
}

\begin{document}

\maketitle

\begin{abstract}
 We present a cross-correlation between the ELAIS 
 15$\mu$m ISO survey with the {\it ROSAT} (0.1-2 keV) all-sky survey and the pointed 
 observations WGACAT source catalogue. 
 The resulting sample contains 15 objects.  
 Optical spectroscopic identifications exist 
for 13 objects: 6 broad-line QSOs, 4 narrow line galaxies or type-2 AGN (NLG)  
 and 3 stars. We have used both the X-ray to IR luminosity flux ratio $f_{x}/f_{IR}$ and 
 the X-ray hardness ratios diagnostics to estimate the amount of obscuration in these objects.    
 The X-ray spectrum of the narrow-line galaxies 
 does not present strong evidence for obscuration; 
 however, the low $f_{x}/f_{IR}$ ratio 
 combined with the high X-ray luminosities 
 suggest that at least one of the narrow line galaxies is associated with 
  an obscured Seyfert nucleus. 4 out of 6 QSOs present high $f_{x}/f_{IR}$
 ratios and steep X-ray spectra with $\Gamma>2$. 
 One QSO (ELAISC15-J133442+375736) at a redshift of z=1.89, 
 has an abnormally low X-ray/IR flux ratio, with its 
 infrared luminosity approaching that of an hyperluminous galaxy 
 ($\sim10^{12.98} h^{-2} L_\odot$).
 Finally, one radio-loud QSO  is the hardest X-ray source in our sample, 
 presenting strong evidence for a high absorbing column ($N_H\sim10^{22}\rm cm^{-2}$).     

\end{abstract}

\vspace{0.2cm}

\section{Introduction}

The standard Unification model  (Antonucci \& Miller 1985) 
 asserts that the nuclei in both type-1 and type-2 Active Galactic Nuclei (AGN) 
 have basically the same structure containing 
 a supermassive black hole, an accretion disk,
 a molecular torus and a broad line region. 
 Then their classification as type-1 or type-2 AGN 
 depends solely on the viewing angle. 
 Specifically, if the source is observed at sufficiently 
 high inclination angle and thus the line of sight 
 intersects the torus, it would be classified as 
 a Seyfert-2, whereas for all other orientations it would
 be deemed to be  a Seyfert-1. X-ray observations of 
 Seyfert-2 galaxies with {\it Ginga}, {\it ASCA} 
 and {\it RXTE} (Smith \& Done 1996; Turner et al. 
 1997; Georgantopoulos \& Papadakis 2001) 
 observe large absorbing columns supporting this scenario. 
 However, optical surveys for AGN have failed so far to produce 
 large numbers of obscured AGN beyond the local Universe.  

In contrast, recent X-ray surveys have proved very useful 
 in finding evidence for the presence
of such obscured AGN  population at moderate to high redshifts ($z\ge 0.1$).
In particular, {\it ROSAT} surveys (0.1-2 keV) have detected a number of 
narrow-line galaxies (NLG) the majority of which are associated with 
obscured AGN (Boyle et al. 1995, Schmidt et al. 1998, Lehmann et al. 2000). 
{\it ASCA} and {\it BeppoSAX} surveys have also found some examples of such an   
obscured AGN population (Boyle et al. 1998; Georgantopoulos et al. 1999; 
Fiore et al. 1999; Akiyama et al. 2000), in the hard 2-10 keV band which is less
prone to photoelectric absorption. These AGN present column densities typically 
higher than $10^{23} \rm cm^{-2}$ while in some cases their optical spectra 
 may present broad lines with only moderate optical reddening. 
 The above results have been corroborated by 
 recent {\it Chandra} and {\it XMM} surveys (eg Mushotzky et al. 2000,
 Brandt et al. 2001, Hasinger et al. 2001). Moreover, 
 {\it ASCA} (Nakanishi et al. 2000), {\it Chandra} (Norman et al. 2001)
 and XMM (Lehmann et al. 2001) have produced the first  examples 
 of the long sought type-2 
 QSO population i.e. with no signs of broad emission lines, similar to 
 Seyfert-2 galaxies in the local Universe.
 
Obscured AGN should emit copious amounts of IR radiation as the 
obscuring matter reprocesses the optical radiation.  
Therefore, the combination of X-ray and IR observations is a powerful 
tool in the detection of such obscured objects. In particular, 
using both the $f_{x}/f_{IR}$ and the X-ray hardness ratios, 
 we can detect the obscured AGN, due to the fact 
that the latter quantities are excellent indicators of high 
photoelectric absorption.
 
Green et al. (1992), combined far-infrared (IRAS) and X-ray (Einstein) 
data, finding a significant correlation between luminosities in 
 the 60-$\mu$m and 0.5-4.5 keV. 
 They also find that the $f_{x}/f_{IR}$ ratios of broad-line AGN 
 are significantly higher  than those of  narrow-line AGN 
 and star-forming galaxies. In a similar way,  
 Boller et al. (1992) cross-correlated the {\it ROSAT} with the IRAS 
 all-sky surveys. The resulting sample consists of 
about 200 objects of which many are obscured Seyfert galaxies 
(Moran, Halpern \& Helfand 1996).
 More recently, Gunn et al. (2001) have 
observed with ISOPHOT onboard ISO 15 sources (QSOs and narrow-line AGN) 
detected in deep {\it ROSAT} fields. The large number of 
narrow line objects detected suggests that large amounts of obscuring matter are 
present in these objects.
Finally, Alexander et al. (2001) observed with {\it BeppoSAX} an 
 area of 0.7 $\rm deg^{2}$ from 
 the ELAIS ISO survey (Oliver et al. 2000, Serjeant et al. 2000)
 to a flux limit of $10^{-13}$ $\rm erg~cm^{-2}~s^{-1}$ in the 2-10 keV band. 
 They find 17 common sources in the hard 2-10 keV band and the 
 $15 \mu m$ band.  Surprisingly, no obscured AGN are among these sources. 

Here we cross-correlate the ELAIS (Oliver et al. 2000; Serjeant et al. 2000) 
15$\mu$m ISO survey with the
{\it ROSAT} all-sky survey and the pointed observations WGACAT source catalogues.
 Willott et al. (2001)  present preliminary results on {\it Chandra} 0.3-10 keV 
observations of ACIS-I fields, covering 0.14 $\rm deg^{2}$ and probing fluxes 
as faint as $\sim10^{-16} \rm erg$ $\rm cm^{-2}$ $\rm s^{-1}$ in the 2-10 keV band.
The {\it ROSAT} observations presented here are complementary as they 
cover the {\it full} area of the ELAIS survey, albeit at much brighter fluxes.
Our aim is to find a number of obscured AGN especially at high redshift
and explore their nature. We note that although the  
{\it ROSAT} passband is soft (0.1-2 keV) and therefore more susceptible to 
photoelectric absorption as compared to either {\it BeppoSAX} or {\it Chandra},
at high redshifts the K-correction diminishes the {\it effective}
obscuring column as it moves the photoelectric absorption 
 cut-off towards lower energies.

The plan of this paper is the following:
In section 2, we describe the ELAIS and {\it ROSAT} catalogues 
 used as well as the cross-correlation results, while in sections 3
 and 4 we present the discussion and conclusions. 

\section{Observations}

\subsection{The European Large Area ISO Survey: (ELAIS)}

The European Large Area ISO Survey (ELAIS) has surveyed $\sim$12 square 
deegres of the sky at 15$\mu$m, 90$\mu$m and subsets of this area at 
6.75$\mu$m and 175$\mu$m using the ISOCAM (Cesarsky et al. 1996) onboard the 
Infrared Space Observatory (ISO), (Kessler et al. 1996).  
The catalogue, a reliable subset of the preliminary analysis 
catalogue (Serjeant et al. 2000), contains 484 sources down to a 
flux limit of $\sim4\rm mJy$ in the 15 $\mu m$ band.
The positional error of the 
$15$ $\mu m$ catalogue is typically 3 arcsec (Serjeant et al. 2000).    
A large number of the 15$\mu$m sources 
(with optical counterparts down to $R\sim 20.5$) have been
spectroscopically identified containing a  
large fraction of AGN and star-forming galaxies 
(Cabrera-Guerra et al., P\'erez-Fournon et al.,
Gruppioni et al. in preparation and Gonz\'alez-Solares et al. 2001).

\subsection{{\it ROSAT} data}
We have used  {\it ROSAT} data from  the PSPC (Position
Sensitive Proportional Catalogue Counter) operating in the 0.1-2.4 keV band. 
In particular, we have used the RASSBSC, RASSFSC and  WGACAT 
point source catalogues. 
RASSBSC ({\it ROSAT} All-Sky Survey Bright Source Catalogue) is derived from 
the all-sky {\it ROSAT} survey and contains 18811 sources (Voges et al. 1999) 
in the energy band 0.1-2.4 keV. 
 RASSFSC ({\it ROSAT} All-Sky Survey Faint Source Catalogue) is an extension 
of the RASSBSC (Voges et al. 2000)
and contains approximately 106,000 sources. 
WGACAT is a point source catalogue generated from all ROSAT PSPC pointed observations 
(see White, Giommi \& Angelini 1994).
The last version of this catalogue 
contains about 88,000 detections, with more than 84,000 individual sources,
obtained from 4160 sequences. 

 The PSPC has an energy 
resolution of 0.5 keV at 1 keV, but limited spatial resolution 
(FWHM $\sim 30 \rm arcsec$).
The positional rms error of the PSPC  detector
is typically  15 arcsec, 
 although the exact error depends on 
the brightness of the source. 
 Throughout this paper, 
 X-ray luminosities are calculated from X-ray fluxes
for a Hubble constant of $H_{\circ}=65$km s$^{-1}$ Mpc$^{-1}$ and 
$\Omega_{\circ}=1$, assuming an average spectrum of 
 $\Gamma=2$.

\subsection{The ELAIS/{\it ROSAT} compilation}
 We have performed a positional cross-correlation of the ELAIS $15\mu m$ 
 catalogue with the above four {\it ROSAT} catalogues, 
 within $\delta \theta \le 0.5 \rm arcmin$. The resulting sample contains 
 15 objects. Note that ELAISC15-J050226-304113 and ELAISC15-J05228-304140
are associated with the same X-ray source; 
 both ELAIS sources are located at the same redshift (z=0.191).  
 We have performed 20 simulations in order to assess the 
 probability of chance coincidences. 
 We are offseting the actual ELAIS source coordinates 
 by a few arcminutes and then we are repeating the 
 cross-correlation. We find 
 that the chance coincidence probability is low. 
 In particular we find that for the RASSBSC we expect 
 $<0.05$ false coincidences. For the RASSFSC 
 and WGACAT the numbers are 0.3 and 0.7 respectively. 
 In Fig. 1 we plot the IR flux distribution 
 of our sources relative to the full ELAIS sample.  
For 13 of our sources, spectroscopic identifications 
exist either through the ELAIS spectroscopic follow-up program 
or through the literature. 
Details of the 15 sources are presented in table 1: 
column (1) ELAIS name; 
column (2) X-ray position; (3) angular separation 
between the IR and X-ray position  
(4) the 15$\mu$m flux (mJy) as listed in 
Serjeant el al. (2000) 
(5) radio flux (mJy) as listed in 
Ciliegi el al. (1999) 
(6) redshift  (7) object classification from: 
 a) Gruppioni et al. (in preparation) b) Cabrera-Guerra et al. (in 
 preparation) or P\'erez-Fournon et al. (in preparation) 
  c) from McHardy et al. (1998) d) NASA Extragalactic Database (NED) 
 (8) the {\it ROSAT} catalogues in which the source has been detected
 ( 1 RASSBSC,2 RASSFSC, 3 WGACAT, 4 ROSHRI).  
 Note that three of our sources have been also detected in the {\it ROSAT} HRI 
 source catalogue, ROSHRI,  (ledas-www.star.le.ac.uk/rosat/rra/roshri) 
 and therefore have more accurate X-ray positions. 
 Hence, we list here the HRI X-ray positions for these sources.   
 In table 2 we present the X-ray properties of our sources: 
 column (2) gives the logarithmic X-ray Luminosity ($\rm erg~s^{-1}$) in the 0.1-2 keV band; 
 (3) the X-ray flux (0.1-2) keV, in units of 10$^{-13}$erg sec$^{-1}$ 
cm$^{-2}$; note that a few of our objects have been detected in more 
than one {\it ROSAT} catalogue; for these we present here the X-ray
flux from the observation with the highest photon statistics.
 (4) X-ray photon index together with the $1\sigma$ error. 
The photon indices were estimated from the hardness ratio 
defined as (h-m)/(h+m) where h and m correspond to the 
0.9-2 keV  and 0.5-0.9 keV (RASSFSC and RASSBSC) 
and 0.9-2 keV and 0.4-0.9 keV bands (WGACAT).
Unfortunately, two objects have  very poor photon statistics and 
therefore their spectral index could not be calculated (ELAISC15-J003015-430333 
and ELAISC15-J003515-433355). 

\begin{figure}
\label{histo}
\mbox{\epsfxsize=7cm \epsffile{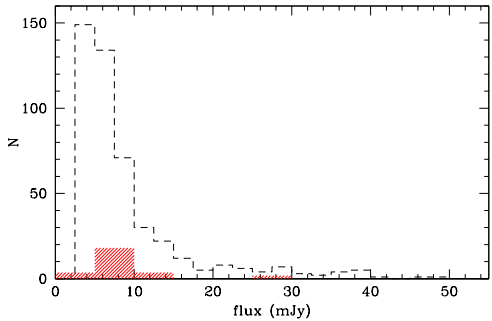}}
\caption{The flux distribution of our 15 sources 
 relative to the ELAIS $15\mu$m flux distribution.} 
\end{figure}

\begin{table*}
\label{ir}
\caption[]{The ELAIS/ROSAT cross-correlation.}
\tabcolsep 6pt
\begin{tabular}{cccccccc} 

Name & X-ray position   & $\delta \theta$ (arcmin)        &$f_{IR}$& $f_{rad}$&$z$ & Classification&   {\it ROSAT} catalogue \\\hline
ELAISC15-J143143+330131 &14 31 43.4 +33 01 31& 0.083	  & 4.95&-& -& 			        & 1,3   \\
ELAISC15-J133451+374616 &13 34 51.3 +37 46 19& 0.078	  &12.13&-&-  & Star$^{d}$			& 1,3,4 \\
ELAISC15-J160623+540555 &16 06 24.1 +54 05 59&0.174	  & 8.00 & 168&0.875& QSO$^{b}$ 		& 2     \\
ELAISC15-J160706+550335 &16 07 04.9 +55 03 58& 0.414	  & 6.03& -&- & Star$^{b}$ 				& 2      \\
ELAISC15-J161521+543147 &16 15 20.5 +54 32 12& 0.423	  & 5.74&-&0.474& QSO$^{b}$		& 2,3     \\
ELAISC15-J163310+405644 &16 33 08.9 +40 56 32&0.289	  & 6.26&-&0.136&NLG$^{b}$ 	        & 2     \\
ELAISC15-J163709+414030 &16 37 08.9 +41 40 54&0.401	  &12.51&8.37&0.765& QSO$^{b, d}$			& 2,3    \\ 
ELAISC15-J003015-430333 &00 30 14.5 -43 03 23&0.190	  &6.15&-&1.564& QSO$^{a}$			& 2     \\
ELAISC15-J003515-433355 &00 35 15.9 -43 33 50&0.183	  & 29.40&-& - &  				& 2     \\
ELAISC15-J050212-302828 &05 02 12.0 -30 28 23&0.083	  &9.38&-&0.86&QSO$^{a}$			& 2,3   \\ 
ELAISC15-J050226-304113 &05 02 26.4 -30 41 28&0.264	  & 9.64&-&0.191&NLG$^{a}$			& 3     \\   
ELAISC15-J050228-304140 &05 02 26.4 -30 41 28&0.398	  &5.02&-&0.191&  NLG			& 3     \\  
ELAISC15-J133401+374912 &13 33 59.6 +37 49 12& 0.28	  & 7.77&-&0.062& NLG$^{d}$			& 3,4   \\
ELAISC15-J133414+375133 &13 34 14.4 +37 51 34& 0.08  & 9.58&-&- & Star$^{c}$			& 3,4   \\  
ELAISC15-J133442+375736 &13 34 44.9 +37 57 17&0.500	  & 4.13&-& 1.89& QSO$^{c}$			& 3    \\ 

\end{tabular}
\end{table*}

\begin{table*}
\caption[]{The X-ray Properties}
\tabcolsep 6pt
\begin{tabular}{cccc} 

Name & $\log L_{x}$     &  $f_x$                                   &  $\Gamma$ \\  
     & $\rm erg~s^{-1}$ & $\rm \times 10^{-13}erg~cm^{-2}~s^{-1}$  &           \\ \hline 
ELAISC15-J143143+330131 & 	 &9.80     &3.17$\pm 0.15$\\                 
ELAISC15-J133451+374616 & 	 &9.95       &2.22 $\pm 0.07$\\              
ELAISC15-J160623+540555 & 44.91  &3.15     &0.43 $\pm 0.78$\\                
ELAISC15-J160706+550335 & 	 &1.35     & 2.84 $\pm 0.81$\\               
ELAISC15-J161521+543147 &44.44	 &4.07     &2.62 $\pm 0.41$\\                
ELAISC15-J163310+405644 &42.65	 &0.90     &0.67 $\pm 1.52$\\                
ELAISC15-J163709+414030 &45.11	 &6.77     &2.84 $\pm 0.22$\\                
ELAISC15-J003015-430333 &45.44	 &3.00       & - \\                          
ELAISC15-J003515-433355 & 	 &5.28     &  -\\                            
ELAISC15-J050212-302828 &44.98	 &3.90     & 2.51 $\pm 0.33$\\               
ELAISC15-J050226-304113 &43.13	 &1.37     &1.85 $\pm 0.48$\\                
ELAISC15-J050228-304140 &43.13	 &1.37     &1.85 $\pm 0.48$\\                
ELAISC15-J133401+374912 &41.28	 &0.19     &2.11 $\pm 0.48$\\                
ELAISC15-J133414+375133 & 	 &0.16     & 3.95 $\pm 0.48$\\               
ELAISC15-J133442+375736 &44.32	 &0.15     & 3.28 $\pm 0.48$\\     
\end{tabular}				 
\end{table*}

\subsection{Notes on Individual Objects}

\begin{itemize}
\item ELAISC15-J160623+540555  
It is the hardest object in X-rays, 
having  $\Gamma=0.43 \pm 0.78$.   
Recently, P\'erez-Fournon et al. (in preparation) discovered that this
object presents broad-lines. 
 The redshift  as measured from the Mg II line is 0.875. 
 If we assume that $\Gamma=1.9$ then
 the column density is  
 $N_H\sim 10^{22}\rm cm^{-2}$ at the 
 QSO's rest-frame.
It is also detected in the 
radio follow-up of the ELAIS fields with flux of 167mJy
(Ciliegi et al. 1999). 
 Furthermore, its $f_x/f_{IR}$ ratio ${\rm log}(f_{x}/f_{IR})=-5.17$ 
is comparable to those of QSOs.

\item $\rm ELAISC15-J143143+330131$.
 This unidentified source presents a  steep X-ray 
spectrum $\Gamma=3.17 \pm 0.15$ together with a  high $f_x/f_{IR}$
 ratio, ${\rm log}(f_{x}/f_{IR})=-4.47$. 
 The above suggest that most probably this source is a QSO.

\item $\rm ELAISC15-J160706+550335$
 Again for this source we have no optical identification. 
 The  ${\rm log}(f_{x}/f_{IR})=-5.41$ ratio for this object 
 places it marginally in 
the narrow-line galaxy regime. Still, it has a steep spectrum with
$\Gamma=2.84 \pm 0.81$. If this object is associated with an
 obscured AGN, the steep X-ray emission may be due to 
 a scattered component. 

\item $\rm ELAISC15-J133442+375736$ 
 This QSO (z=1.89) presents a  very steep spectrum in X-rays 
($\Gamma=3.28 \pm 0.48$) with
X-ray luminosity of $L_{x}=2\times10^{44}$erg s$^{-1}$. 
 This object could be associated with  
an intrinsically weak X-ray source. 
Interestingly, the infrared luminosity  
is very high $L_{IR} \sim 10^{12.98} h^{-2}L_{\odot}$
approaching the luminosity of a hyperluminous galaxy.

\item $\rm ELAISC15-J163310+405644$:
 This is classified as a star-forming galaxy 
 according to P\'erez-Fournon et al., (in preparation).
 Its luminosity is $L_{x}=4\times10^{42}$ erg s$^{-1}$, 
 higher than those of known star-forming galaxies. 
 The high X-ray luminosity is in stark contrast 
 with the optical spectrum. 
 This object bears close resemblance 
 with the ``composite'' Seyfert/star-forming galaxies
 of Moran et al. (1996). 
 Unfortunately, the X-ray spectrum has poor photon statistics 
 and thus it cannot provide additional constraints on the nature of
 this object.  The ${\rm log}(f_{x}/f_{IR})=-5.6$ is low typical of 
 obscured Seyfert galaxies in the local Universe.  
 Finally, considering that this source is an AGN then 
 the estimated infrared luminosity,  
$L_{IR} \simeq 2\times 10^{11} h^{-2}L_{\odot}$ respectively.

\item $\rm ELAISC15-160706+550335$
This is classified as a star according to 
P\'erez-Fournon et al., (in preparation).
The optical magnitude is very bright ($R \sim 12$). 

\item $\rm ELAISC15-J133451+374616$
This is classified as a star according to the 
NASA Extragalactic Database (NED). 
Still its $f_{x}/f_{IR}$ ratio is high for a normal star
(see Stocke et al. 1991) suggesting that 
this is probably an X-ray binary.   

\end{itemize}

\section{DISCUSSION}

In figure 2 we plot the estimated ${\rm log}(f_{x}/f_{IR})$ as a 
function of redshift for our 15 sources.
The lines model the expected evolution of the $f_x/f_{IR}$ 
ratio for various classes of extragalactic object, 
namely QSOs, Seyfert-2 and narrow-line galaxies. The solid line (QSOs) 
is based on the average QSO SED (spectral energy distribution) 
produced by Elvis et al. (1994). 
The dash line (star-forming galaxies) and the dot-dash line (Seyfert-2) 
are based on infrared SEDs generated using the Xu et al. (1998) model
(see also Alexander et al. 2001). 
The errors on the above lines correspond to 
the statistical spread in X-ray/IR colours. 
Note that, the 
QSO X-ray/IR ratios in our sample are in general in good agreement with the 
corresponding $f_x/f_{IR}$ model. 
 The only exception is ELAISC15-J133442+375736, 
the highest redshift QSO (z=1.89) in our sample.  
The narrow line galaxies are again in good agreement with 
the predicted $f_{x}/f_{IR}$ ratio with the possible 
 exception of ELAISC15-J050228-304140 which has a somewhat higher 
 $f_x/f_{IR}$ approaching the QSO regime. 
 Note however that this is the ``double'' ISO source at z=0.191
 which is associated with a single X-ray source. 
 Therefore we are unable   
 to assess the exact amount of the X-ray emission which 
 originates from each galaxy and to draw conclusions 
 on the nature of this object based on the $f_x/f_{IR}$ ratio alone. 

\begin{figure}
\label{ratio}
\mbox{\epsfxsize=9cm \epsffile{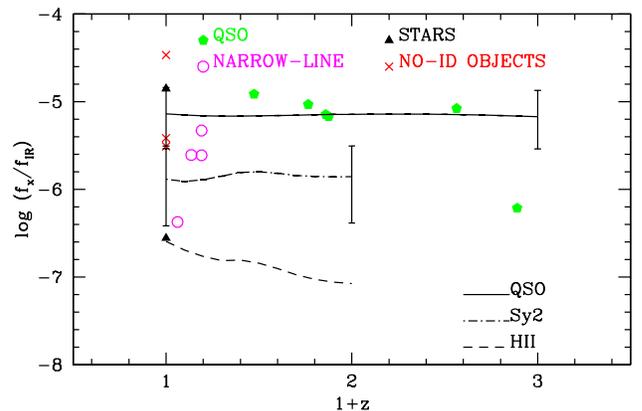}}
\caption{The X-ray/IR flux ratios for various types of galaxies as a 
function of $1+z$. The lines denote the predicted ratios (see Elvis et al. 
1994; Alexander et al. 2001), while the errors correspond to the statistical spread in X-ray/IR 
colours for source type.} 
\end{figure}

\begin{figure}
\label{gamma}
\mbox{\epsfxsize=9cm \epsffile{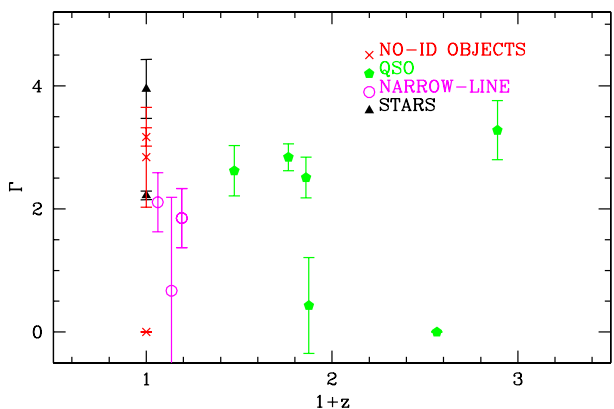}}
\caption{The X-ray spectrum as a function of $1+z$.} 
\end{figure}

The photon indices provide additional evidence on the nature of 
 these objects. In Fig. \ref{gamma} we present the derived 
X-ray spectrum, $\Gamma$, as a function of redshift
 while in Fig. 4  the 
 ${\rm log} (f_{x}/f_{IR})$ ratio as a function of $\Gamma$. 
 From Fig. 4 it becomes obvious that most QSOs have soft X-ray spectra
 ($\Gamma>2$). These values are quite typical of QSO spectra in the soft 
 {\it ROSAT} band (Schartel et al. 1996; Fiore et al. 1997; Blair et al. 2000).
Interestingly, the high redshift QSO ELAISC15-J133442+375736 which presents the lowest 
 $f_x/f_{IR}$ ratio, compared to the other QSOs,
 also shows a steep spectrum.
  The steep spectral index observed ($\Gamma\approx3.3\pm0.5$)
 clearly argues against an absorbed AGN scenario.
 Therefore, the low $f_{x}/f_{IR}$ ratio could be 
 attributed to either an intrinsically low 
 X-ray emission (similar to PG1011-040, Gallagher et al. 2001) 
 or alternatively to high IR luminosity.
 Indeed, this source has a luminosity of $\sim10^{12.98} h^{-2} L_\odot$.      
 For comparison we note that  our source has an $f_{x}/f_{IR}$ ratio a few times lower 
 than that of the hyperluminous source ($\sim1.2 \times 10^{13} h^{-2} L_\odot$) 
 ELAISP90-J164010+410502 (Morel et al. 2001) which  
 has been detected in the 90$\mu m$ ELAIS sample. 
  The hardest X-ray source in our sample is 
 ELAISC15-J160623+540555. This is a radio-loud QSO at a redshift of 
 z=0.875. The hardness ratio corresponds 
 to a photon index of 0.43 or alternatively to a column 
 density of $N_H\sim10^{22} \rm cm^{-2}$ (assuming 
 $\Gamma=1.9)$) at the QSO's rest-frame. This is not unphysical for
 radio-loud QSOs. Indeed,  
 Reeves \& Turner (2000) using {\it ASCA} observations of a sample 
 of 35 radio-loud QSOS, find that a large fraction 
 of these present absorbing columns of the order of 
 $10^{22} \rm cm^{-2}$ or higher.  
 The same conclusions have been reached by Fiore et al. (1998) 
 using {\it ROSAT}.

The narrow-line galaxies have soft X-ray spectra all consistent with 
 $\Gamma\sim 2$. Only  ELAISC15-J163310+405644 presents a harder X-ray spectrum
 ($\Gamma=0.67\pm1.52)$ but within the large uncertainty
 this is still consistent with the spectra of the other narrow-line objects.  
 Such soft X-ray spectra are compatible with X-ray observations of 
 star-forming galaxies in the {\it ROSAT} band 
 (see Read et al. 1997; Dahlem et al. 1998). 
 The X-ray luminosities of these objects  yield additional clues 
 on the nature of these objects. It is becoming clear that 
 star-forming galaxies cannot exceed luminosities of 
 $10^{42}$ $\rm erg~s^{-1}$ in the X-ray band 
 (eg Moran et al. 1996; Zezas, Georgantopoulos \& Ward 1998). 
 Then the high luminosity of ELAISC15-J163310+405644 implies the 
 presence of an active nucleus in this object. 
 Still, the optical spectrum presents no sign 
 of nuclear activity. This object has very 
 similar properties to the ``composite'' galaxies 
 presented in detail in Veron et al. (1997) 
 and Moran et al.  (1996). The ``composites'' have  
 optical spectra which would classify these
 as star-forming galaxies on the basis of the 
 diagnostic line ratio diagram of Veilleux \& Osterbrock (1987),
 but still their X-ray luminosities are typical of AGN. 
  The ``double'' ISO source again presents 
 a high X-ray luminosity  ($\sim 10^{43}\rm erg~s^{-1}$).
 Then both or at least one of the sources are associated with an 
 AGN. Alternatively, the X-ray source could be associated with 
 a cluster of galaxies (indeed both ELAIS sources 
 have the same redshift) while the IR emission 
 could originate  from star-forming processes in the two galaxies.
  Finally, ELAISC15-J133401+374912 presents a low X-ray luminosity 
 typical of star-forming galaxies suggesting that 
 this may be the only ``bona-fide'' normal galaxy in our sample. 
 Of course the possibility that this hosts a low luminosity AGN cannot be 
 ruled out. 

\begin{figure}
\label{gratio}
\mbox{\epsfxsize=9cm \epsffile{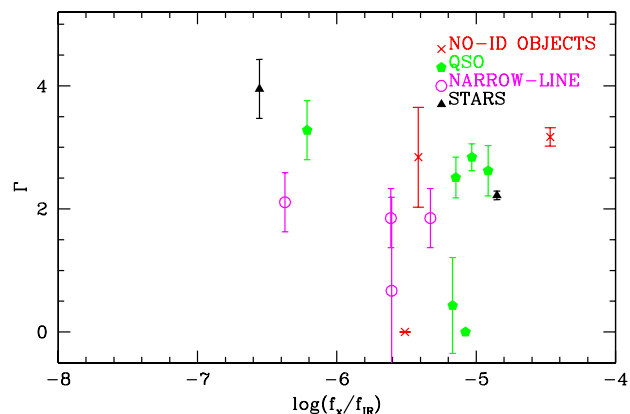}}
\caption{The X-ray spectrum versus ${\rm log}(f_{x}/f_{IR})$ }
\end{figure}


\section{Summary and Conclusions}

We have presented a cross-correlation between the ELAIS (Oliver et al. 2000; Serjeant et al. 2000) 
15$\mu$m ISO survey with the {\it ROSAT} (0.1-2 keV) all-sky survey and pointed 
observations source catalogues. We found  
15 common objects, 13 of which have optical indentifications: 
6 broad-line QSOs ($z=0.47-1.89$), 4 narrow-line galaxies or type-2 AGN 
 ($z=0.06-0.19)$ and 3 stars.
Utilising the measured $f_{x}/f_{IR}$ and the X-ray hardness ratios 
 we can investigate whether our sources present large amounts of obscuration. 
 We find that 5/6 QSOs present steep spectra arguing 
 against the presence of large (more than a few times $\rm 10^{21}~cm^{-2}$) 
 obscuring columns.
 The hardest source in our sample is a radio-loud QSO 
  having a column density of $N_H\sim10^{22}$. 
 This is a radio-loud QSO and therefore 
 the presence of a large obscuring column is not surprising. 
  The highest redshift QSO has a low $f_x/f_{IR}$ ratio
  suggesting that its X-ray emission 
  is  weak  relative to its 
 powerful IR luminosity which 
 is borderline to that of hyperluminous galaxies. 

On the other hand, the narrow-line galaxies in general have 
soft X-ray spectra being consistent with 
$\Gamma\sim 2$ and therefore 
 with both  the star-forming galaxy and AGN spectra. 
 ELAISC15-J163310+405644 presents a harder X-ray spectrum
($\Gamma=0.67\pm1.52)$ but within the large uncertainty
this is still consistent with the spectra of the other narrow-line objects.    
 On the basis of the high X-ray luminosity ($>10^{42}$ $\rm erg~s^{-1}$), 
 it becomes  clear that at least one of the four narrow line objects 
 (ELAISC15-J163310+405644) harbours an AGN. Then the low X-ray to IR flux ratio 
 implies that we are only viewing the scattered or 
 star-forming component in the {\it ROSAT} passband. 
 This object presents additional interest as the optical spectrum classifies 
 it as a star-forming galaxy and demonstrates that 
 X-ray observations play a critical role
 in identifying the true nature of IR sources.  

 In conclusion,  the majority 
 of extragalactic objects in our sample falls roughly 
 into two categories: high redshift non-obscured 
 radio-quiet QSOs and  
 low redshift absorbed AGN or star-forming galaxies. 
 We have not found any evidence for the presence of 
 absorbed radio-quiet QSOs at high redshift
 similar to those found in {\it BeppoSAX} or 
 {\it ASCA} surveys. Deeper X-ray observations 
 of a few of the ELAIS fields 
 with {\it Chandra} and {\it XMM} are expected  to 
 shed more light on the number density 
 and the nature of these enigmatic objects.

\section* {Acknowledgements}
This work was supported by EC Network programme 'POE' (grant number HPRN-CT-2000-00138).
IPF, FCG and EGS were supported by project PB98-0409-CO2-01 of the 
Spanish Ministerio de Ciencia y Techolog\'ia.


{\small 
\begin{thebibliography}{}
\bibitem[]{}Akiyama, M., et al., 2000, ApJ, 532, 700
\bibitem[]{}Alexander, M. D., et al., 2001, ApJ, 554, 18
\bibitem[]{}Antonucci, R., \& Miller, J. S., 1985, ApJ, 297, 621
\bibitem[]{}Blair, A. J., Stewart, G. C., Georgantopoulos, I., Boyle, B. J.,
Griffiths, R. E., Shanks, T., Almaini, O., 2000, MNRAS, 314, 138
\bibitem[]{}Boller, T., Meurs, E. J. A., Brinkmann, W., Fink, H.,
Zimmermann, U., Adorf, H. M., 1992, A\&A, 261, 57
\bibitem[]{}Boyle, B. J., McMahon, R.G., Wilkes, B.J., Elvis, M., 1995, MNRAS, 272, 462
\bibitem[]{}Boyle, B. J., Almaini, O., Georgantopoulos, I., 
Blair, A. J., Stewart, G. C., Griffiths, R. E., Shanks, T., Gunn, K. F., 1998, MNRAS, 297, L53
\bibitem[]{}Brandt, W. N., Hornschemeier, A. E., Schneider, D. P., 
Alexander, D. M., Bauer, F. E., Garmire, G. P., Vignali, C., 2001, ApJ, 558, L5
\bibitem[]{}Cesarsky, C. J., et al., 1996, A\&A, 315, L32
\bibitem[]{}Ciliegi, P., et al., 1999, MNRAS, 302, 222
\bibitem[]{}Dahlem, M., Weaver, K. A., Heckman, T. M., 1998, ApJS, 118, 401
\bibitem[]{}Gallagher, S. C., Brandt, W. N., Sambruna, R. M., Mathur, S., Yamasaki, N., 2001, ApJ, 546, 795
\bibitem[]{}Georgantopoulos, I., Almaini, O., Shanks, T., Stewart, G. C.
Griffiths, R. E., Boyle, B. J., Gunn, K. F., 1999, MNRAS, 305, 125
\bibitem[]{}Georgantopoulos, I., Papadakis, I. E., 2001, MNRAS, 305, 125
\bibitem[]{}Gonz\'alez-Solares et al., 2001, in Proc. of ESA Conference, 
''The Promise of First'', in press
\bibitem[]{}Green, P. J., Anderson, S. F., Ward, M. J., 1992, MNRAS, 254, 30
\bibitem[]{}Gunn, K. F., et al., 2001, MNRAS, submitted
\bibitem[]{}Elvis, M. et al., 1994, ApJS, 72, 1
\bibitem[]{}Fiore, F., Matt, G., Nicastro, F., 1997, MNRAS, 284, 731
\bibitem[]{}Fiore, F., Elvis, M., Giommi, P., Padovani, P., 1998, ApJ, 492, 79
\bibitem[]{}Fiore, F., La Franca, F., Giommi, P., Elvis, M., Matt, G., Comastri, A.,
Molendi, S., Gioia, I., 1999, MNRAS, 306, L55
\bibitem[]{}Hasinger, G., et al., 2001, A\&A, 365, 45
\bibitem[]{}Kessler, M. F., et al., 1996, A\&A, 315, L27
\bibitem[]{}Lehmann, I., et al. 2000, A\&A, 354, 35
\bibitem[]{}Lehmann, I., Hasinger, G., Murray, S. S., Schmidt, M., 2001,
Proceedings for X-rays at Sharp Focus Chandra Science Symposium
\bibitem[]{}McHardy, I. M., et al., 1998, MNRAS, 295, 641
\bibitem[]{}Morel, T., et al., 2001, MNRAS, 327, 1187
\bibitem[]{}Moran, E. C., Halpern, J. P., Helfand, D. J., 1996, ApJS, 106, 341
\bibitem[]{}Mushotzky, R. F., Cowie, L. L., Barger, A. J., Arnaud, K. A., 2000,
Nature, 404, 459
\bibitem[]{} Nakanishi, K., Akiyama, M., Ohta, K., Yamada, T., 2000, ApJ, 534, 587 
\bibitem[]{} Norman, C., et al., 2001, submitted, {\it astro-ph/0103198}
\bibitem[]{}Oliver, S., et al., 2000, MNRAS, 316, 749
\bibitem[]{}Read, A. M., Ponman, T. J., Strickland, D. K., 1997, MNRAS, 286, 629
\bibitem[]{}Reeves, J. N., Turner, M. J. L., 2000, MNRAS, 316, 234
\bibitem[]{}Schartel, N., et al., 1996, MNRAS, 283, 1015
\bibitem[]{}Schmidt, M., et al., 1998, A\&A, 329, 495
\bibitem[]{}Serjeant, S., et al., 2000, MNRAS, 316, 768
\bibitem[]{}Smith, D. A., \&, Done, C., 1996, MNRAS, 280, 355
\bibitem[]{}Stocke, J. T., Case, J., Donahue, M., Shull, J. M., Snow, T. P., 1991, ApJ, 374, 72
\bibitem[]{}Turner, T. J., George, I. M., Nandra, K., Mushotzky, R. F., 1997, ApJ, 488, 164
\bibitem[]{}White, N. E., Giommi, P., Angelini, L., 1994, AAS, 185, 4111
\bibitem[]{}Veilleux, S.; Osterbrock, D. E., 1987, ApSJ, 63, 295
\bibitem[]{}Veron, P., Goncalves, A. C., Veron-Cetty, M.-P., 1997, A\&A, 319, 52
\bibitem[]{}Voges, W., et al., 1999, A\&A, 349, 389
\bibitem[]{}Voges, W., et al., 2000, IAUC, 7432, 3
\bibitem[]{} Willott, C.J. et al., 2001, astro-ph/0105560  
\bibitem[]{}Xu, C., et al., 1998, ApJ, 508, 576
\bibitem[]{}Zezas, A. L., Georgantopoulos, I., Ward, M. J., 1998, MNRAS, 301, 915

\end{thebibliography}
}
\end{document}